\documentstyle[aps,epsfig,amssymb]{revtex}

\begin{document}

\title{On the theory of superconductivity in  ferromagnetic 
superconductors with triplet pairing}
\author{V.P. Mineev and T. Champel}
\address{
Commissariat \`{a} l'Energie Atomique,  DSM/DRFMC/SPSMS\\
17 rue des Martyrs, 38054 Grenoble Cedex 9, France}
\date{\today}

\maketitle

\begin{abstract}
We point out that ferromagnetic superconductors with triplet pairing 
and strong spin-orbit coupling are even in the simplest case at least 
two-band superconductors. 
The Gor'kov type formalism for such superconductors is developed
and the Ginzburg-Landau equations are derived.  The dependence of the critical
temperature on the concentration of ordinary point-like impurities is
found.  Its nonuniversality could serve as a qualitative measure of
the two-band character of ferromagnetic superconductors.  The problem
of the upper critical field determination is also discussed.

\end{abstract}


\section{Introduction}

The extension of the Bardeen-Cooper-Schrieffer (BCS) theory of
superconductivity to the case with two bands of itinerant
electrons was developed soon after the appearance of the BCS theory
\cite{1}.  This theory has been the subject of renewed interest
following the recent discovery of $\mathrm{MgB}_{2}•$ - the first
superconducting material where the existence of two energy gaps has
been unambiguously demonstrated by thermodynamic and spectroscopic
measurements \cite{2}.  Certainly, there are many other
superconducting compounds where multiband effects are less pronounced
and experimentally invisible because the Cooper pairing occurs mostly
in one band of the itinerant electrons or holes.  On the other hand,
there is a whole class of superconductors where two-band (or more
generally multiple band) superconductivity is an inherent property:
the so called ferromagnetic superconductors where the different bands
with spin "up" and spin "down" electrons are always present. 
$\mathrm{UGe}_{2}•$ \cite{3,4}, $\mathrm{ZrZn}_{2}•$
\cite{5}, and $\mathrm{URhGe}$ \cite{6} are recent examples of such
materials where the superconducting states are expected to be spin
triplet in order to avoid the large depairing influence of the
exchange field due to ferromagnetism.
    
The symmetry classification of the superconducting states for 
itinerant
ferromagnetic spin-triplet superconductors has been proposed recently by
several authors \cite{7,8,9}.  At the same time, a general
Gor'kov-type mathematical description of the multi-band superconductivity
in a ferromagnetic metal with a triplet pairing has not been developed yet.
The principal goal of this article is to present such a description for a two-band
ferromagnetic metal with an anisotropic spectrum of quasiparticles and a
general form of the pairing interaction.  Being useful for a concrete
calculation with a particular form of the spectrum and the pairing interaction,
this approach allows to solve in general terms several problems typical  for the 
superconductivity theory, such as: the critical temperature
determination, the derivation of the Ginzburg-Landau equations, the suppression of the
superconductivity by impurities, the upper critical field calculation.

We begin with the general form of the order parameter
and the pairing interaction in a two-band itinerant ferromagnet. 
Then the Gor'kov equations will be written. That permits to calculate
the spectrum of the quasiparticles and the critical temperature, and to
derive the system of the coupled equations for the order parameters from
two bands.  Then, a law of suppression of the critical temperature by
point-like nonmagnetic impurities is found.  Its characteristic
nonuniversal behavior can serve as a qualitative measure of the
two-band character of ferromagnetic superconductors.  The problem of the 
upper critical field determination is finally discussed.

\section{Ferromagnetic superconductors with triplet pairing}

\subsection{Two-band superconductivity}

For a triplet superconductor the order parameter
is written as \cite{10}
\begin{eqnarray}
\Delta_{\alpha \beta}({\bf R},{\bf k})=
\left(
\begin{array}{cc}
\Delta_{\uparrow}¥ & \Delta_{0}¥ \\
\Delta_{0}¥ &\Delta_{\downarrow}¥ \end{array}\right) &=&\left({\bf
d}^{\Gamma}({\bf R},{\bf k})
\bbox{\sigma}\right)i\sigma_{y}\nonumber\\
&=&\left(
\begin{array}{cc}
-d_{x}({\bf R},{\bf k})+i d_{y}({\bf R},{\bf k})& d_{z}({\bf R},{\bf 
k}) \\
d_{z}({\bf R},{\bf k}) & d_{x}({\bf R},{\bf k})+i d_{y}({\bf R},{\bf 
k})
\end{array}
\right),
\label{e1}
\end{eqnarray}
where  $\bbox{\sigma}=(\sigma_{x},\sigma_{y},\sigma_{z})$ are the 
Pauli matrices.
The superconducting states ${\bf d}^{\Gamma}({\bf 
R},{\bf k})$ with different critical temperatures in the
ferromagnetic crystals are classified in accordance with the irreducible
co-representations $\Gamma$ of the magnetic group $M$ of the
crystal~\cite{7,8,9}.  All the co-representations in ferromagnets with
orthorombic and cubic symmetries are one-dimensional.  However, they
obey multicomponent order parameters determined through the
coordinate dependent pairing amplitudes: one per each band populated
by electrons with spins "up" or "down" and one per each pair of the
bands with the opposite spins (zero spin projection states).  Owing to
the big difference in the Fermi momenta, the pairing of electrons
from the different bands is negligibly small.  Hence, we shall neglect
the pairing amplitude with zero spin projection; in another words
$\Delta_{0}¥=d_{z}({\bf R},{\bf k})=0$ will be taken throughout the
paper.  Also, we limit ourself to the consideration of two band
ferromagnetic superconductors with a strong spin-orbital coupling, when the
two-component order parameter has the form
\begin{equation}
{\bf d}^{\Gamma}({\bf R},{\bf k})=\frac{1}{2}
[-(\hat{x}+i\hat{y})\Delta_{\uparrow}¥({\bf R},{\bf k})+
(\hat{x}-i\hat{y})\Delta_{\downarrow}¥({\bf R},{\bf k})]
\label{e2}
\end{equation}
which was first
pointed out in the paper \cite{11}.  Here $\hat{x}$, $\hat{y}$  are the
unit vectors of the spin (or, more exactly, of the pseudospin \cite{10})
coordinate system pinned to the crystal axes.
\begin{equation} 
\Delta_{\uparrow}¥({\bf R},{\bf k})=-\eta_{1}¥({\bf R})f_{-}¥({\bf
k}),~~~~ \Delta_{\downarrow}¥({\bf R},{\bf k})=\eta_{2}¥({\bf
R})f_{+}¥({\bf k}).
\label{e3}
\end{equation}
The functions $f_{\pm}¥({\bf k})=f_{x}¥({\bf k})\pm
if_{y}¥({\bf k})$ and the projections $f_{i}({\bf k})$, with $i=x,y$ are
odd functions of the momentum directions of the pairing particles on the Fermi
surface.   The general forms of these functions for the
different co-representations in ferromagnetic superconductors with
orthorhombic and cubic symmetries are  listed in the paper~\cite{7}.
For instance, in the case of the $A_{1}¥$ representation in orthorombic
crystal, they are
\begin{equation}
f_{x}({\bf k})=
k_{x}¥u_{1}¥^{A_{1}¥}¥+ik_{y}¥u_{2}¥^{A_{1}¥}¥, ~~ f_{y}({\bf k})=
k_{y}¥u_{3}¥^{A_{1}¥}¥+ik_{x}¥u_{4}¥^{A_{1}¥}¥,
\label{e4}
\end{equation}
where $u_{1}¥^{A_{1}¥}¥,\ldots $ are real functions of $k_{x}¥^{2}¥, 
k_{y}¥^{2}¥, k_{z}¥^{2}¥$.
The simple consequence of this is that the 
nodes in the quasiparticle spectrum of the superconducting A-states in
orthorombic ferromagnets dictated only by the symmetry are the nodes lying on the northern and southern
poles of the Fermi surface $k_{x}¥=k_{y}¥=0$.  On the contrary, for the
B-states they are on the line of the equator $k_{z}¥=0$.

The  order parameter amplitudes
$\eta_{1}¥({\bf R})$ and $\eta_{2}¥({\bf R})$ (which are coordinate dependent and complex)  have been considered (and discussed) in
the paper\cite{7} as being equal. This is in general not true. However, even in the
general case, they are not completely independent:
\begin{equation}
\eta_{1}¥({\bf R})=|\eta_{1}¥({\bf R})|e^{i\varphi({\bf R})}¥,
~~~\eta_{2}¥({\bf R})=|\eta_{2}¥({\bf R})| e^{i{\varphi({\bf
R})}¥\pm i\pi}.
\label{e5}
\end{equation}
Although being different by their modulous, they have the same phase with an
accuracy $\pm \pi$.  The latter property guarantees the consistency of the
transformation of both parts of the order parameter under the time
reversal.

The BCS Hamiltonian in a two-band ferromagnet with a triplet pairing is
\begin{equation}
H=\sum_{{\bf k},{\bf k}',\alpha}\langle {\bf k} |\hat{h}_{\alpha}¥|{\bf
k}'\rangle a^{\dag}_{ {\bf k}\alpha} a_{{\bf k}'\alpha} +\frac{1}{2} 
\sum_{{\bf k},{\bf k}', {\bf q},\alpha,\beta} V_{\alpha
\beta}({\bf k},{\bf k}') a^{\dag}_{
-{\bf k}+{\bf q}/2, \alpha} a^{\dag}_{ {\bf k} +{\bf q}/2,\alpha} a_{
{\bf k}' +{\bf q}/2,\beta} a_{ -{\bf k}'+{\bf q}/2, \beta},
\label{e6}
\end{equation}
where the band indices $\alpha$ and $\beta$ are $(\uparrow,
\downarrow)$ or $(1,2)$,
\begin{equation}
\hat{h}_{\alpha}¥=\hat{\varepsilon}_{\alpha}¥ 
-\mu_{B}¥\hat{\bf g}_{\alpha}¥{\bf H}_{\mathrm{ext}}¥/2 +U({\bf
r})-\varepsilon_{F}
\label{e7}
\end{equation}
are one particle band energy operators, the
functions $\hat{\varepsilon}_{\alpha}¥ $ (including the exchange splitting)
and the $\hat{\bf g}_{\alpha}¥$-factor depend on the gauge invariant
operator $-i \nabla +({e}/{c}) {\bf A}({\bf r})$ and on the crystallographic
directions.  In the simplest case of isotropic bands without a spin-orbital
coupling, ${\bf g}_{1,2}¥=\pm 2{\bf H}_{\mathrm{ext}}¥/H_{\mathrm{ext}}¥$.  $U({\bf r})$
is an impurity potential, ${\bf H}_{\mathrm{ext}}$ is an external
magnetic field,
\begin{equation}
\nabla \times {\bf A} = {\bf B}={\bf H}_{\mathrm{ext}}+4
\pi{\bf M},
\label{e8}
\end{equation} 
and ${\bf M}$ is the magnetic moment of the ferromagnet.

The pairing potential interaction is expanded as
\begin{equation}
V_{\alpha\beta}({\bf k},{\bf k}') =-V_{\alpha\beta}\varphi_{\alpha}¥
({\bf k})\varphi_{\beta}¥^{*}¥({\bf k'}),
\label{e9}
\end{equation}
where
\begin{equation}
\varphi_{\uparrow}({\bf k})=-f_{-}({\bf k}),~~~~
\varphi_{\downarrow}({\bf k})=f_{+}({\bf k}).
\label{e10}
\end{equation}
It contains four different interaction terms which can be typically classified into two groups corresponding to: (i) a pairing
between electrons with the same spin polarization (intraband
interaction) and (ii) the interband scattering terms with 
$V_{\uparrow\downarrow}¥=V_{\downarrow\uparrow}¥$ describing the transitions of
the pair electron from one sheet of the Fermi surface to the other
sheet by reversing the pair spin orientation with the help of the
spin-orbit coupling.

When the interband scattering is negligible $V_{\uparrow\downarrow}¥=
V_{\downarrow\uparrow}¥=0$, the pairing of the electrons occurs first
only in one of the sheets of the Fermi surface like in the $A_{1}$
phase of $^{3}$He.  In general, the superconductivity in each band is
not independent.

\subsection{Gor'kov equations}

We want to determine  the Green's functions of ferromagnetic 
superconductors in the absence of external perturbations and impurity 
scattering.
Even under these simple conditions, the system is not spatially 
uniform
due to the inherent presence of $ 4\pi M$.
If we neglect $4 \pi M$,  the system is spatially uniform.  
Then, we can 
write the Gor'kov equations in the form
\begin{eqnarray}
\left(i\omega_{n}-\xi_{{\bf k}\alpha}\right) G_{\alpha}({\bf k},\omega_{n})+
\Delta_{\alpha}({\bf k}) F_{\alpha}^{\dagger}({\bf k},\omega_{n})=1
\nonumber \\
\left(i\omega_{n}+\xi_{{\bf k}\alpha}\right) F_{\alpha}^{\dagger}¥ ({\bf
k},\omega_{n})+\Delta_{\alpha}^{\dagger}¥ ({\bf k}) G_{\alpha}({\bf
k},\omega_{n})=0
\label{e11}
\end{eqnarray}
where $\xi_{{\bf k}\alpha}=\varepsilon_{{\bf k}\alpha}-\varepsilon_{F}$
and $\omega_{n}¥=\pi T(2n+1)$ are Matsubara frequencies.  The equations for
each band are only coupled through the order parameter given by the
self-consistency condition
\begin{equation}
\Delta_{\alpha}({\bf k})=-T\sum_{n}\sum_{{\bf
k}'}\sum_{\beta=\uparrow,\downarrow} V_{\alpha,\beta}\left( {\bf k},{\bf
k}'\right) F_{\beta}({\bf k}',\omega_{n}).
\label{e12}
\end{equation}
The superconductor Green's functions are
\begin{eqnarray}
G_{\alpha}\left({\bf k},\omega_{n}\right) &=& -\frac{i\omega_{n}+\xi_{{\bf
k}\alpha}}{\omega_{n}^{2}+E_{{\bf k},\alpha}^{2}} \\
F_{\alpha}\left({\bf k},\omega_{n}\right)&=& \frac{\Delta_{\alpha} ({\bf
k})}{ \omega_{n}^{2}+E_{{\bf k},\alpha}^{2} },
\label{e13}
\end{eqnarray}
where
$E_{{\bf k},\alpha}=\sqrt{\xi_{{\bf
k}\alpha}^{2}+\left|\Delta_{\alpha}({\bf k}) \right|^{2}} $. 
Obviously, the superconductivity in ferromagnetic superconductors is
non-unitary. 

The Gor'kov equations taking into consideration the
magnetic moment $4 \pi M$, an external field and non-magnetic
point-like impurities can be easily written according to the general
procedure described in \cite{10}.  We shall not overload the paper by
this and just write the self-consistency equations near the
superconducting transition.

\subsection{The order parameter equations near the superconducting 
transition}

This system consists of two equations for the order parameter 
components with spin polarizations "up" and "down",

\begin{eqnarray}
\Delta_{\alpha}\left({\bf R},{\bf r}\right)&=& -T \sum_{n,\beta}
\int d {\bf r}' V_{\alpha, \beta}\left({\bf r},{\bf r}'\right)
G_{\beta}¥({\bf r}',\tilde{\omega}_{n}¥^{\beta}¥) G_{\beta}¥({\bf
r}',-\tilde{\omega}_{n}¥^{\beta}¥) \nonumber \\
&&\times
\exp
\left[
i {\bf r}' {\bf D}({\bf R})
\right]
\left\{
\Delta_{\beta} \left({\bf R},{\bf r}'\right) +
\Sigma_{\beta}(\tilde{\omega}_{n}^{\beta},{\bf R}) \right\},
\label{e14}
\end{eqnarray}
and two equations for the impurity self-energy components
\begin{eqnarray}
\Sigma_{\alpha}\left(\tilde{\omega}_{n}^{\alpha},{\bf R}\right)&=&
n_{\mathrm{i}} u^{2}_{\alpha} \int d {\bf r}\, G_{\alpha}¥({\bf
r},\tilde{\omega}_{n}¥^{\alpha}¥) G_{\alpha}¥({\bf
r},-\tilde{\omega}_{n}¥^{\alpha}¥) \exp\left[ i {\bf r} {\bf D}({\bf
R}) \right] \nonumber \\
&&\times
\left\{
\Delta_{\alpha}\left({\bf R},{\bf r}\right) +
\Sigma_{\alpha}(\tilde{\omega}_{n}^{\alpha},{\bf R}) \right\}
\label{e15},
\end{eqnarray}
where
 $\tilde{\omega}_{n}^{\alpha}=\omega_{n}+{\mathrm sign} \, \omega_{n}/2
 \tau_{\alpha}$, and $\tau_{\alpha}$ is the quasi-particle mean free time in
 the different bands.
These mean free times are related in the Born approximation to the 
impurity concentration $n_{\mathrm{i}}$ through
\begin{equation}
\frac{1}{2 \tau_{\alpha}}= \pi n_{\mathrm{i}} N_{0\alpha}u^{2}_{\alpha} ,
\end{equation}
with $u_{\alpha}$ the amplitude of the impurity scattering, and
$N_{0\alpha}$ the density of electronic states in each band.

The operator of covariant differentiation is
$${\bf D}({\bf
R})=-i\frac{\partial}{\partial {\bf R}}+\frac{2e}{c}
{\bf A}({\bf R}).$$
The normal metal electron Green functions
are
\begin{equation}
G_{\alpha}¥({\bf r},\tilde{\omega}_{n}¥^{\alpha}¥) =\int \!  \frac{d
{\bf p}}{(2\pi)^{3}} e^{i {\bf p} \cdot {\bf r}}
\left(i\tilde{\omega}_{n}^{\alpha}-\xi_{{\bf p},\alpha}
+\mu_{B}{\bf g}_{{\bf p},\alpha}
{\bf H}_{\mathrm{ext}}/2\right)^{-1}.
\label{e16}
\end{equation}
The order parameter components in different bands are determined in
accordance with (\ref{e3}):
\begin{equation}
\Delta_{\uparrow}¥({\bf R},{\bf r})=
-\eta_{1}¥({\bf R})f_{-}¥({\bf r}),~~~~ \Delta_{\downarrow}¥({\bf
R},{\bf r})= \eta_{2}¥({\bf R})f_{+}¥({\bf r}).
\label{e17}
\end{equation}

\subsection{The critical temperature $T_{c0}$}

In the absence of an external field let us first find the critical temperature
$T_{c0}¥$ of a pure ferromagnetic superconductor in the (formally) spacially uniform situation of a negligible
ferromagnetic moment ${\bf M}=0$.  In this case the anomalous impurity
self-energy part $\Sigma_{\alpha}(\widetilde{\omega}_{n}^{\alpha},{\bf
R})=0$, and from (\ref{e14}) we obtain the system of equations
\begin{eqnarray}
\eta_{1}¥ &=&(g_{1}¥\eta_{1}¥+
g_{12}¥\eta_{2}¥)\lambda(T_{c0}¥),\nonumber\\
\eta_{2}¥&=&(g_{21}¥\eta_{1}¥+
g_{2}¥\eta_{2}¥)\lambda(T_{c0}¥),
\label{e18}
\end{eqnarray}
where 
$g_{1}¥=
V_{\uparrow \uparrow}¥\langle|f_{-}¥({\bf k})|^{2}¥ N_{0\uparrow }¥
(\hat{\bf k})\rangle $, the angular brackets mean the
averaging over the Fermi surface, $N_{0\uparrow }¥(\hat{\bf k})$ is
the angular dependent density of electronic states at the Fermi
surface of the band $\uparrow$.  Correspondingly
$g_{12}¥=
V_{\uparrow \downarrow}¥\langle|f_{+}¥({\bf k})|^{2}¥ N_{0\downarrow }¥
(\hat{\bf k})\rangle $, $g_{21}¥=
V_{\downarrow\uparrow
}¥\langle|f_{-}¥({\bf k})|^{2}¥ N_{0\uparrow }¥ (\hat{\bf k})\rangle 
$, and $g_{2}¥=V_{\downarrow \downarrow}¥\langle|f_{+}¥({\bf k})|^{2}¥ N_{0\downarrow }¥(\hat{\bf
k})\rangle $. 
The function $\lambda(T)$ is 
\begin{equation}
\lambda(T)=2\pi T\sum_{n\ge 0}¥\frac{1}{\omega_{n}¥}=\ln
\frac{2\gamma \epsilon}{\pi T}~,
\label{e19}
\end{equation}
$\ln\gamma=0,577\ldots $ is the Euler constant, $\epsilon $ is an
energy cutoff.
 
Thus, similar to \cite{1} the critical temperature is given by
\begin{equation}
T_{c0}¥=(2\gamma\epsilon/\pi)\exp{(-1/g)},
\label{e20}
\end{equation}
where $g$ is defined by the maximum value for which the determinant of the system is zero
(\ref{e18})
\begin{equation}
g= (g_{1}¥+g_{2}¥)/{2} +\sqrt{({g_{1}¥-g_{2}¥})^{2}¥/{4}+g_{12}¥g_{21}¥}.
\label{e21}
\end{equation}
In particular, at $g_{12}¥,~g_{21}¥~\ll~g_{1}¥,~g_{2}¥$, the critical
temperature is determined by
\begin{equation}
g= \max(g_{1}¥,~g_{2}¥).
\label{e22}
\end{equation}

\subsection{The critical temperature dependence on impurities concentration}

Triplet superconductivity is
suppressed by non-magnetic impurities \cite{12}.  Moreover, the law of
suppression of the superconductivity is described by the universal
Abrikosov-Gor'kov (AG) dependence \cite{13}
\begin{equation}
    -\ln t=\Psi\left( \frac{1}{2}+ \frac{x}{4\gamma t}\right)
-\Psi\left( \frac{1}{2}
\right)
\label{e23}
\end{equation} 
valid for any unconventional superconducting state and applicable 
in particular to a concrete unconventional superconductor 
independently of the
pressure \cite{10}. Here $\Psi$ is the digamma function. The 
variable $t=T_{c}•/T_{c0}•$ is the ratio
of the critical temperature of the superconductor with a given
concentration of impurities $n_{\mathrm{i}}$ to the critical 
temperature of the clean
superconductor, and $x 
=n_{\mathrm{i}}/n_{\mathrm{ic}}•=\tau_{\mathrm{c}}•/\tau$ is the 
ratio 
of the
impurity concentration in the superconductor to the critical 
impurity concentration destroying
superconductivity, or the inverse ratio of the corresponding mean 
free particle lifetimes.  The critical mean free time is given by 
$\tau_{\mathrm{c}}•=\gamma/\pi T_{c0}• $.  This dependence has been
demonstrated (although with some dispersion of the experimental
points) for the triplet superconductor
$\mathrm{Sr}_{2}•\mathrm{RuO}_{4}•$ \cite{14}.
    
Deviations from the universality of the
AG law can be caused by the anisotropy of the scattering which
takes place in the presence of extended imperfections in the 
crystal.  Such
a modification of the theory applied to
$\mathrm{UPt}_{3}•$ has been considered previously \cite{15}.  
However, a
complete experimental investigation of the suppression of
superconductivity by impurities in this unconventional 
superconductor, in particular the study of the universality of the 
behavior, has not been performed.

The nonuniversality of the suppression of
superconductivity can also be caused by any inelastic   
scattering mechanism by impurities with internal degrees of
freedom of magnetic or nonmagnetic origin.  For the simplest
discussion of this, see \cite{16}. 

Finally, universality
is certainly not expected in multiband superconductors.  Theories 
for this case have been developed with regard to the unconventional
superconductivity in $\mathrm{Sr}_{2}•\mathrm{RuO}_{4}•$ (p-wave, 
two-band
two-dimensional model \cite{17}) and conventional 
superconductivity in
$\mathrm{MgB}_{2}•$ (anisotropic scattering two-band model 
\cite{18}).
    
A simple modification of the universal AG law
for the suppression of the superconductivity by impurities in a
two-band ferromagnetic superconductor is derived here.  Our consideration
is limited to the simplest case of scattering by ordinary point-like
impurities.  Then, due to spin conservation, one can neglect interband
quasi-particle scattering and take into account only the intraband
quasi-particle scattering on impurities.  
At finite impurity concentration the 
system of equations similar to (\ref{e18}) is:
\begin{eqnarray}
\eta_{1}¥&=&g_{1}¥\Lambda_{1}¥(T)\eta_{1}¥+
g_{12}¥\Lambda_{2}¥(T)\eta_{2}¥,\nonumber\\
\eta_{2}¥&=&g_{21}¥\Lambda_{1}¥(T)\eta_{1}¥+ g_{2}¥\Lambda_{2}¥(T)\eta_{2}¥,
\label{e24}
\end{eqnarray}
where
\begin{equation}
\Lambda_{1,2}¥(T)= \Psi\left(\frac{1}{2}\right)-\Psi\left(\frac{1}{2}+
\frac{1}{4\pi\tau_{1,2}¥T}\right)+\ln\frac{T_{c0}¥}{T}+
\lambda(T_{c0}¥).
\label{e25}
\end{equation}

Hence, the critical temperature is determined from the equation
\begin{equation}
\left(g_{1}¥\Lambda_{1}¥(T)-1\right) \left(g_{2}¥\Lambda_{2}¥(T)-1\right)-
g_{12}¥g_{21}¥\Lambda_{1}¥(T)\Lambda_{2}¥(T)=0.
\label{e26}
\end{equation}
In particular, at $g_{12}¥,~g_{21}¥~\ll~g_{1}¥,~g_{2}¥$ the critical temperature
is determined by the $\max(T_{c1}¥,T_{c2}¥)$ of the solutions of the equations
\begin{equation}
\ln\frac{T_{c0}¥}{T_{c1}¥}=\Psi\left(\frac{1}{2}+
\frac{1}{4\pi\tau_{1}¥T_{c1}¥}\right)-\Psi\left(\frac{1}{2}\right)+
\frac{1}{g_{1}¥}-\lambda(T_{c0}¥),
\label{e27}
\end{equation}
\begin{equation}
\ln\frac{T_{c0}¥}{T_{c2}¥}=\Psi\left(\frac{1}{2}+
\frac{1}{4\pi\tau_{2}¥T_{c2}¥}\right)-\Psi\left(\frac{1}{2}\right)+
\frac{1}{g_{2}¥}-\lambda(T_{c0}¥).
\label{e28}
\end{equation}
Let us accept for determination that $g_{1}¥> g_{2}¥$. Hence, the  maximal
critical temperature in the absence of impurities is defined by
$1/g_{1}¥=\lambda(T_{c0}¥)$.  Then, at small impurity concentrations the
solutions of (\ref{e27}) and (\ref{e28}) are the linear functions of impurities
concentration:
\begin{equation}
T_{c1}¥=T_{c0}¥~-~\frac{\pi}{8\tau_{1}¥},
\label{e29}
\end{equation}
\begin{equation}
T_{c2}¥=T_{c0}¥-\frac{1}{g_{2}¥}+\frac{1}{g_{1}¥}~-~\frac{\pi}{8\tau_{2}¥}.
\label{e30}
\end{equation}
These lines can in principle intersect each other. As a result, an upturn
in the critical temperature dependence on impurity concentration
$T_{c}¥(n_{i}¥)$ appears.
Such a type of  deviation of the $T_{c}¥(n_{i}¥)$ dependence from
the AG-law presents the direct manifestation of the two-band character of the
superconductivity.  On the other hand, an absence of strong deviations
from the universal one-band curve if found experimentally in
a ferromagnetic superconductor would mean that the superconductivity is
developed in one-band with only electrons with "up" spins paired and
the "down" spin electrons leave normal (or vice versa).

Another specific feature of the ferromagnetic superconductors is that even
in the absence of an external magnetic field the exchange field
$H_{\mathrm{ex}}•\sim { E_{\mathrm{ex}}•/\mu_{B}•}$ acting on the
electron spins in a ferromagnet produces in addition an
electromagnetic field $4\pi M \sim 4 \pi {\mu_{ B}•k_{F}•^{3}•} $
acting via the electronic charges on the orbital motion of electrons,
and suppressing the superconductivity~\cite{footnote}.  Hence, the
actual critical temperature in ferromagnetic superconductors is always
smaller by the value $\sim 4 \pi M •/H_{c2}•(T=0)$ relative to the
(imaginary) ferromagnetic superconductor without $4 \pi M•$.  The
upper critical field $H_{c2}•$ is also purity dependent.  That is why
the impurity concentration dependence of the actual $T_{c}•$ in a
ferromagnetic superconductor might be determined not only directly by
the suppression of superconducting correlations by the impurity
scattering as in any nonconventional superconductor but also
indirectly through the supression of $H_{c2}•$.  In fact the second
indirect mechanism has a negligible influence because the ratio $4 \pi
M•/H_{c2}•(T=0)$ is of the order of $10^{-2}•$ for superconductors
with an upper critical field of the order of several Teslas.

Thus, the problem of the determination of the critical temperature in a
superconducting ferromagnet is at bottom the problem of the determination
of the upper critical field in a single domain ferromagnet.

\subsection{The upper critical field}

The equations for the determination of the upper critical field at least near
$T_{c}¥$ are easily derived from the system (\ref{e14})-(\ref{e15}).
Keeping only the lowest order gradient terms we have
\begin{eqnarray}
\Delta_{\alpha}\left({\bf R},{\bf r}\right)&=& -T \sum_{n,\beta}
\int d {\bf r}' V_{\alpha, \beta}\left({\bf r},{\bf r}'\right)
G_{\beta}¥({\bf r}',\tilde{\omega}_{n}¥^{\beta}¥) G_{\beta}¥({\bf
r}',-\tilde{\omega}_{n}¥^{\beta}¥) \nonumber \\
&&\times
\left\{\left(1- \left({\bf r}' {\bf D}({\bf R}) \right)^{2}¥/2\right) 
\Delta_{\beta} \left({\bf R},{\bf r}'\right) +
\left(i{\bf r}' {\bf D}({\bf R}) \right)
\Sigma_{\beta}(\tilde{\omega}_{n}^{\beta},{\bf R}) \right\},
\label{e31}
\end{eqnarray}
and 
\begin{eqnarray}
\Sigma_{\alpha}\left(\tilde{\omega}_{n}^{\alpha},{\bf R}\right)&=&
n_{\mathrm{i}} u^{2}_{\alpha} \int d {\bf r}\, G_{\alpha}¥({\bf
r},\tilde{\omega}_{n}¥^{\alpha}¥) G_{\alpha}¥({\bf
r},-\tilde{\omega}_{n}¥^{\alpha}¥) \nonumber \\
&&\times\left\{\left( i {\bf r} {\bf D}({\bf R}) \right)
\Delta_{\alpha}\left({\bf R},{\bf r}\right) +
\Sigma_{\alpha}(\tilde{\omega}_{n}^{\alpha},{\bf R}) \right\}
\label{e32}.
\end{eqnarray}
Finding $\Sigma_{\alpha}(\tilde{\omega}_{n}^{\alpha},{\bf R})$ from
the last equation and substituting it into (\ref{e31}), we obtain after all
the necessary integrations the pair of the Ginzburg-Landau equations
for the two components of the order parameter
\begin{eqnarray}
\eta_{1}¥&=&V_{\uparrow\uparrow}¥\hat\alpha_{1}¥\eta_{1}¥+
V_{\uparrow\downarrow}¥\hat\alpha_{2}\eta_{2}¥,\nonumber\\
\eta_{2}¥&=&V_{\uparrow\downarrow}¥\hat\alpha_{1}\eta_{1}¥+
V_{\downarrow\downarrow}¥\hat\alpha_{2}\eta_{2}¥,
\label{e33}
\end{eqnarray}
where the operator $\hat\alpha_{1}$ consists of the previously determined homogeneous part and of the 
second order gradient terms
\begin{equation}
\hat\alpha_{1}¥=\langle|f_{-}¥({\bf k})|^{2}¥ N_{0\uparrow }¥
(\hat{\bf k})\rangle\Lambda_{1}¥(T)-K_{\uparrow ij}¥D_{i}¥D_{j}¥.
\label{e34}
\end{equation}
The gradient terms coefficients are
\begin{eqnarray}
K_{\uparrow ij}¥&=&\langle|f_{-}¥({\bf k})|^{2}¥ N_{0\uparrow }¥ (\hat{\bf
k})v_{F\uparrow i}¥(\hat{\bf k})v_{F\uparrow j}¥(\hat{\bf k})\rangle
\frac{\pi T}{2}\sum_{n\ge
0}¥\frac{1}{|\tilde\omega^{\uparrow}¥_{n}¥|^{3}¥}\nonumber\\
&+&\langle
f_{-}¥({\bf k}) N_{0\uparrow }¥ (\hat{\bf k})v_{F\uparrow i}¥(\hat{\bf
k})\rangle\langle f_{-}¥^{*}¥({\bf k}) N_{0\uparrow }¥ (\hat{\bf
k})v_{F\uparrow j}¥(\hat{\bf k})\rangle \frac{\pi^{2}¥T
n_{i}¥u_{\uparrow}¥^{2}¥}{2} \sum_{n\ge
0}¥\frac{1}{\omega_{n}¥^{2}¥\tilde\omega^{\uparrow}¥_{n}¥^{2}¥}
\label{e35}.
\end{eqnarray}
The operator $\hat\alpha_{2}$ is obtained from (\ref{e34})-(\ref{e35}) by the natural
substitutions $1\to 2$, $\uparrow~ \to~ \downarrow$, $+~\to ~- $.

Now, the problem of the upper critical field finding is just the
resolution of the two coupled equations (\ref{e33}).  There are a
lot of different situations depending on the crystal symmetry, the direction
of spontaneous magnetization and on the external field orientation.  The
simplest case is when the external magnetic field is parallel or
antiparallel to the easy magnetization axis.  If this latter coincides
with the 4-th order symmetry axis in the cubic crystal as it is the case in
$\mathrm{ZrZn}_{2}¥$, then the gradient terms in the perpendicular plane are
isotropic and described by the two constants $K_{\uparrow
ij}¥=K_{\uparrow}¥\delta_{ij}¥$ and $K_{\downarrow ij}¥=
K_{\downarrow}¥\delta_{ij}¥$.  This case formally corresponds to the
problem of the determination of the upper critical field parallel to the
c-direction in the two-band hexagonal superconductor $\mathrm{MgB}_{2}¥$ solved in
\cite{20}.  Then, the linearized Ginzburg-Landau equations describe a
system of two coupled oscillators and have their solution in the form
$\eta_{1}¥=c_{1}¥f_{0}¥(x)$ and $\eta_{2}¥=c_{2}¥f_{0}¥(x)$, where
$f_{0}¥(x)=\exp(-hx^{2}¥/2)$, and $h$ is related to the upper critical field 
by means of
\begin{equation} 
|H_{c2}¥\pm 4\pi M|=\frac{h(\tau)\Phi_{0}¥}{2\pi},
\label{e36}
\end{equation}
where $\Phi_{0}¥$ is the flux quantum. 

Let us for simplicity limit
ourself to the impurityless case.  Then, $\tau=1-T/T_{c0}¥$ and the
equation for the determination of the upper critical field is
\begin{eqnarray}
&&\left[g_{1}¥(\tau+
\lambda(T_{c0}¥))+V_{\uparrow\uparrow}¥K_{\uparrow}¥h-1\right]
\left[g_{2}¥(\tau
+\lambda(T_{c0}¥))+V_{\downarrow\downarrow}¥K_{\downarrow}¥h-1\right]\nonumber\\
&-&
\left[g_{12}¥(\tau
+\lambda(T_{c0}¥))+V_{\uparrow\downarrow}¥K_{\uparrow}¥h-1\right]
\left[g_{21}¥(\tau
+\lambda(T_{c0}¥))+V_{\uparrow\downarrow}¥K_{\downarrow}¥h-1\right]=0
\label{e37}.
\end{eqnarray}
This is a simple square equation and, as before, if we consider the case
$g_{12}¥, g_{21}¥\ll g_{1}¥, g_{2}¥$ and $g_{1}¥>g_{2}¥$, then we
obtain the two following roots
\begin{equation}
h_{1}¥(\tau)=\frac{g_{1}¥\tau}{V_{\uparrow\uparrow}¥K_{\uparrow}¥},
\label{e38}
\end{equation}
\begin{equation}
h_{2}¥(\tau)=\frac{g_{2}¥}{V_{\downarrow\downarrow}¥K_{\downarrow}¥}\left(
\tau+\frac{1}{g_{1}¥}-\frac{1}{g_{2}¥}\right)
\label{e39}.
\end{equation}
This two lines can in principle intersect each other. As a result,  an upturn in the 
temperature dependence of the upper critical field appears.

In the most anisotropic situation such as in the orthorombic crystals $\mathrm{UGe}_{2}¥$
and $\mathrm{URhGe}$, all the coefficients
$K_{\uparrow xx}$, $K_{\uparrow yy}¥$, $K_{\downarrow xx}¥$ and
$K_{\downarrow yy}¥$ are different, even for an external field direction parallel or
antiparallel to the easy magnetization axis.  Then, our system of equations can be
solved following a variational approach developed in \cite{20}.  Again,
an upturn in the $h(\tau)$ dependence can be possible.

The comparison with an experiment masked by the
presence of many ferromagnetic domains will be not easy.  Monodomain measurements
are possible in high enough fields.  To work in this region one can
easily obtain the fourth order gradient terms contributing to the Ginzburg-Landau
equations.  However, the problem of the theoretical determination of the
upper critical field at arbitrary temperature has the same
principal difficulty as in any conventional anisotropic superconductor
\cite{21}.

\section{Conclusion}

Ferromagnetic superconductors are in general multiband metals. The 
two-band description of ferromagnetic superconductors with a triplet 
pairing developed in this paper presents the simplest model 
applicable to this type of material. 
One-band superconductivity in these superconductors arises only at a 
negligibly small spin-orbit coupling, as  it is the case for the 
$A_{1}-$phase of $^{3}$He.
We studied the dependence of the critical temperature $T_{c}$
on the 
concentration  of ordinary point-like impurities in the framework of 
a two-band weak coupling BCS theory. 
We demonstrated  that the non-universal $T_{c}(x)$ dependence could
serve as a qualitative measure of the two band character of the
superconductivity in ferromagnetic superconductors.
Also, the general equations for the determination of the upper critical
field at arbitrary temperature and impurity concentration were derived.
The solution of these equations near the critical temperature was found
in the simplest case of a cubic crystalline symmetry for the field
orientation parallel or antiparallel to the 4-th order symmetry axis.

\section{Acknowledgements}
The authors are indebted to Dr. M. Zhitomirskii for the valuable discussions
about the physics of multi-band superconductivity.


\begin{thebibliography}{99}
\bibitem{footnote}
This is correct of course only if $4 \pi M > H_{c1}$ (where 
$H_{c1}$  is  the lower critical field).
\end{thebibliography}
\end{document}